\documentclass[prd,aps,floatfix,twocolumn,showpacs,nofootinbib,preprintnumbers,superscriptaddress]{revtex4-1}

\usepackage{amsmath}
\usepackage{multirow}
\usepackage{mathtools}
\usepackage{graphicx}
\usepackage{amssymb}

\graphicspath{{.}{graphics/}}

\newcommand{\Lagrangian}[0]{\mathcal{L}}

\newcommand{\eqcomma}[0]{\, ,}
\newcommand{\eqstop}[0]{\, .}

\newcommand{\eq}[1]{Eq.~(\ref{eq:#1})}
\newcommand{\fig}[1]{Fig.~\ref{fig:#1}}
\newcommand{\tab}[1]{Table \ref{tab:#1}}
\renewcommand{\sec}[1]{Sec.~\ref{sec:#1}}

\newcommand{\dc}[1]{\mathrm{d}^3 #1 \,}

\newcommand{\bra}[1]{\langle \, #1 \, | \,}
\newcommand{\ket}[1]{\, | \, #1  \, \rangle}
\newcommand{\atlarget}{\xrightarrow{\text{large } t}}

\newcommand{\order}[1]{\mathcal{O}\left(#1\right)}

\renewcommand{\Re}{\operatorname{Re}}
\renewcommand{\Im}{\operatorname{Im}}
\newcommand{\I}{\mathbb{I}}

\begin{document}

\title{Disconnected contributions to the spin of the nucleon}

\author{A.~J.~Chambers}
\affiliation{CSSM, Department of Physics,
 University of Adelaide, Adelaide SA 5005, Australia}
\email{alexander.chambers@adelaide.edu.au}

\author{R.~Horsley}
\affiliation{School of Physics and Astronomy,
 University of Edinburgh, Edinburgh EH9 3JZ, UK}

\author{Y.~Nakamura}
\affiliation{RIKEN Advanced Institute for Computational Science,
 Kobe, Hyogo 650-0047, Japan}

\author{H.~Perlt}
\affiliation{Institut f\"ur Theoretische Physik,
 Universit\"at Leipzig, 04103 Leipzig, Germany}

\author{D.~Pleiter}
\affiliation{JSC, J\"ulich Research Centre,
 52425 J\"ulich, Germany}
\affiliation{Institut f\"ur Theoretische Physik,
 Universit\"at Regensburg, 93040 Regensburg, Germany}

\author{P.~E.~L.~Rakow}
\affiliation{Theoretical Physics Division,
 Department of Mathematical Sciences,
 University of Liverpool, Liverpool L69 3BX, UK}

\author{G.~Schierholz}
\affiliation{Deutsches Elektronen-Synchrotron DESY,
 22603 Hamburg, Germany}

\author{A.~Schiller}
\affiliation{Institut f\"ur Theoretische Physik,
 Universit\"at Leipzig, 04103 Leipzig, Germany}

\author{H.~St\"uben}
\affiliation{Regionales Rechenzentrum,
 Universit\"at Hamburg, 20146 Hamburg, Germany}

\author{R.~D.~Young}
\affiliation{CSSM, Department of Physics,
 University of Adelaide, Adelaide SA 5005, Australia}

\author{J.~M.~Zanotti}
\affiliation{CSSM, Department of Physics,
 University of Adelaide, Adelaide SA 5005, Australia}

\preprint{\vbox{\hbox{
      ADP-15-29/T931, DESY 15-156, Edinburgh 2015/21, LTH 1056
}}}

\begin{abstract}
  The spin decomposition of the proton is a long-standing topic of much
  interest in hadronic physics.
  Lattice QCD has had much success in calculating the connected
  contributions to the quark spin.
  However, complete calculations, which necessarily involve gluonic
  and strange-quark contributions,
  still present some challenges.
  These ``disconnected'' contributions typically involve small signals
  hidden against large statistical backgrounds and rely on
  computationally intensive stochastic techniques.
  In this work we demonstrate how a Feynman-Hellmann approach may be
  used to calculate such quantities, by measuring shifts in the proton
  energy arising from artificial modifications to the QCD action.
  We find a statistically significant non-zero result for the
  disconnected quark spin contribution to the proton of about $-5$\%
  at a pion mass of 470 MeV.
\end{abstract}

\maketitle

\section{Introduction}

The simple quark model picture suggests that the total nucleon spin is
comprised entirely in terms of its constituent quark spins. In
contrast, experimental measurements reveal that the quark spin only
generates about one third of the total nucleon spin
\cite{Alexakhin:2006oza}. This observation is a clear representation
of the nontrivial dynamics associated with nonperturbative
QCD. Resolving the full composition of the nucleon spin in terms of
the QCD degrees of freedom remains an active experimental and
theoretical pursuit.  For an overview of the status and progress, we refer the
reader to
Refs.~\cite{Anselmino:1994gn,Filippone:2001ux,Bass:2004xa,Myhrer:2007cf,Thomas:2008ga,Aidala:2012mv}.

As a systematically improvable method for studying nonperturbative
properties of QCD, lattice simulations offer the potential to provide
quantitative predictions for the decomposition of the nucleon
spin. For recent numerical investigations of the nucleon spin, and
related matrix elements, see
Refs.~\cite{Bratt:2010jn,Syritsyn:2011vk,Dinter:2011sg,Owen:2012ts,Capitani:2012gj,Sternbeck:2012rw,Alexandrou:2013joa,Bhattacharya:2013ehc,Bali:2013nla}.

In the conventional approach, spin matrix elements are extracted from
3-point correlation functions. Operator insertions that are directly
connected to the quark field operators of the nucleon interpolators
can be reliably computed using established techniques. The operator
insertions that involve self-contracted fermion lines, which are
essential to isolate the strangeness spin content, for instance, require the
stochastic estimation of the trace of an all-to-all propagator. Owing
to the increased computational demand of this stochastic estimator and
a relatively weak numerical signal, such disconnected contributions
have often been neglected in lattice simulations. Nevertheless, there
has been substantial progress made in recent years
\cite{Babich:2010at,QCDSF:2011aa,Engelhardt:2012gd,Abdel-Rehim:2013wlz,Deka:2013zha}. For
a related calculation involving the vector current matrix elements we
also refer to Ref.~\cite{Green:2015wqa}.

In recent work, we have proposed an alternative to the
conventional 3-point function technique for the study of hadron matrix
elements in lattice QCD. By adapting the Feynman-Hellmann (FH) theorem to
the lattice framework, we are able to isolate matrix elements in terms
of an energy shift in the presence of an appropriate weak external
field \cite{Horsley:2012pz,Chambers:2014qaa}. This is similar to the
technique proposed by Detmold in Ref.~\cite{Detmold:2004kw}.
In Ref.~\cite{Horsley:2012pz} we used the Feynman-Hellmann relation to
extract the gluonic contribution to the nucleon mass. The application
of Feynman-Hellmann was further developed in
Ref.~\cite{Chambers:2014qaa} for the study of the connected spin
contributions in various hadrons.
We have also recently shown how it is possible to compute
flavour-singlet renormalisation constants nonperturbatively by an
appropriate application of the FH theorem \cite{Chambers:2014pea}.

In the present work, we apply the FH technique to resolve disconnected
spin matrix elements. Whereas the connected spin contributions could
be computed on conventional gauge fields, the disconnected
contributions requires the generation of new special-purpose gauge
configurations. The influence of the weak external spin field is
therefore accumulated through the importance sampling of the hybrid
Monte Carlo simulation. While such new configurations come at
significant computational cost, the computing time is comparable to that
required for reliable with sampling using the conventional stochastic
techniques.

The manuscript proceeds as follows: Section II reviews the
implementation of the FH theorem for the extraction of spin matrix
elements and summarises the lattice simulation parameters;
Section III describes the strategy for the isolation of the relevant
matrix elements from the two-point correlation functions; with results
reported in Section IV; followed by concluding remarks in Section V.

\section{Feynman-Hellmann methods and simulation details}
\label{sec:fh_methods}

Here we discuss the Feynman-Hellmann approach in the context of
calculations of disconnected contributions to matrix elements,
in particular the quark axial charges.
For details of previous calculations of the connected contributions, and
the Feynman-Hellmann technique in general,
see \cite{Chambers:2014qaa}.

The quark axial charges are defined by forward matrix elements of the
axial operator,
\begin{equation}
  \bra{\vec{p}, \vec{s}}
  \bar{q}(0) \gamma_\mu \gamma_5 q(0)
  \ket{\vec{p}, \vec{s}}
  =
  2 i s_\mu \Delta q
  \eqstop
\end{equation}
We access disconnected contributions to these quantities by
implementing a modification to the fermion part of the QCD Lagrangian
during gauge-field generation.
Extra terms are included involving the axial operator weighted by some
freely-varying real parameter $\lambda$, applied equally to all three
quark flavours,
\begin{equation}
  \Lagrangian
  \to
  \Lagrangian
  +
  \lambda \sum_{q = u,d,s} \bar{q} \gamma_3 \gamma_5 q
  \eqstop
  \label{eq:lag_mod}
\end{equation}
This operator satisfies $\gamma_5$-hermiticity,
and so the determinant of the fermion matrix is
still real. Hence we avoid introducing any sign problems.
We choose projection matrices to isolate spin-up and down components of
the nucleon correlation function,
\begin{equation}
  \Gamma_\pm =
  \frac{1}{2}(\I + \gamma_4)
  \frac{1}{2}(\I \pm i \gamma_5 \gamma_3)
  \eqcomma
  \label{eq:proj_mats}
\end{equation}
and by application of the Feynman-Hellmann relation,
find that the correlator picks up a complex phase which mimics an
imaginary energy component,
\begin{equation}
  E \to E(\lambda) + i \phi(\lambda)
  \eqstop
  \label{eq:fh_phase}
\end{equation}
At first order in the parameter $\lambda$, there is no shift in the
real part of the energy, and the shift in the phase is exactly equal to
the disconnected contribution to the total quark axial charge,
\begin{equation}
  \left.
  \frac
  {\partial E}{\partial \lambda}
  \right|_{\lambda=0}
  =
  0
  \qquad
  \qquad
  \left.
  \frac
  {\partial \phi}{\partial \lambda}
  \right|_{\lambda=0}
  =
  \pm
  \Delta \Sigma_\text{disc.}
  \eqcomma
  \label{eq:fh_rel}
\end{equation}
where the total contribution is the sum of the individual flavour
contributions,
\begin{equation}
  \Delta \Sigma_\text{disc.} =
  \Delta u_\text{disc.}
  +
  \Delta d_\text{disc.}
  +
  \Delta s
  \eqstop
\end{equation}
Note that we access the total contribution because the
operator in \eq{lag_mod} includes terms for all 3 quark flavours.
Also note that the strange contribution is purely disconnected.
The different signs in \eq{fh_rel} result from the different choices
of $\Gamma_\pm$, and we note that changing the spin projection is
equivalent to flipping the sign of $\lambda$.

Our strategy for this calculation, motivated by \eq{fh_rel},
is to generate new gauge ensembles for multiple values of
$\lambda$, measure the phase shift in \eq{fh_phase} and determine
$\Delta \Sigma_\text{disc.}$ from the linear behaviour.

In our previous work, we were able to access the connected part by
implementing the change in \eq{lag_mod} to the Dirac matrix before
inversion to compute the quark propagator entering hadron correlation
functions (see \cite{Chambers:2014qaa}). Here the modification is made
to the fermion matrix in the HMC algorithm, and so information about
the purely disconnected contributions is accessed.

\subsection{Simulation details}

We use gauge field configurations with $2+1$ flavours of
non-perturbatively $O(a)$-improved Wilson fermions and a
lattice volume of $L^3 \times T = 32^3 \times 64$.
The lattice spacing $a = 0.074(2)$ fm is set using a number of singlet
quantities \cite{Horsley:2013wqa,Bornyakov:2015eaa,Bietenholz:2010jr,Bietenholz:2011qq}.
The clover action used comprises the tree-level Symanzik improved
gluon action together with a stout smeared fermion action, modified
for the implementation of the Feynman-Hellmann method
\cite{Chambers:2014qaa}.

For the results discussed here, we use ensembles with two sets
of hopping parameters, $(\kappa_l,\kappa_s) =$
(0.120900,120900) and (0.121095,0.120512),
corresponding to pion masses of approximately 470 and 310 MeV.
These have been
generated with the modified quark action described in \eq{lag_mod}.
The details of these ensembles, including the values of $\lambda$
realised, are given in \tab{ensembles}.

\begin{table}
  \begin{tabular}{c l | c c c c}
    \hline
    \hline
    $(\kappa_l, \kappa_s)$ & $\lambda$ & $N_\text{conf.}$ & $N_\text{sources}$ & $\phi$ \\
    \hline
    \multirow{3}{*}{(0.120900, 0.120900)} & -0.0125 & 500 & 1 & 0.0014(10) \\
    & -0.00625 & 500  & 1 & 0.00002(83) \\
    & 0.03    & 500  & 1 & -0.00237(77) \\
    \hline
    \multirow{2}{*}{(0.121095, 0.120512)} & -0.025   & 600  & 1 & -0.0008(13) \\
    & 0.05     & 800  & 5 & 0.00027(61) \\
    \hline
    \hline
  \end{tabular}
  \caption{Table of ensembles generated for this work. Two pion masses
  with three and two values of $\lambda$ respectively have been
  used. The number of configurations and sources used, as well as the
  phase shift measured (discussed in \sec{analysis} and \sec{results})
are also listed.}
  \label{tab:ensembles}
\end{table}

\section{Analysis techniques}
\label{sec:analysis}

A standard zero-momentum projected nucleon correlation function is
given by
\begin{equation}
  G_\pm(t)
  =
  \int \dc{\vec{x}}
  \Gamma_\pm
  \bra{\Omega}
  N(x) \bar{N}(0)
  \ket{\Omega}
  \notag
  \atlarget
  A e^{-Et}
  \eqcomma
  \label{eq:normal_corr}
\end{equation}
where $N$ and $\bar{N}$ are interpolating operators coupling to the
nucleon
ground state, and the projection matrices $\Gamma_\pm$ (defined in
\eq{proj_mats}) project spin-up and down components respectively.
For our simulations, we use identical source and sink smearing and
operators. Hence, the amplitude $A$ is purely real.

With the modification to the Lagrangian in \eq{lag_mod}, an imaginary
component is introduced to the exponential factor, in addition to
a complex shift in the amplitude. This shift in the amplitude is not
the focus of this work, but is related to the
$\lambda$ dependence of the wavefunction overlap factors.
To first order in $\lambda$, the amplitude and energy take
the form
\begin{align}
  A & \to A + \lambda (\Delta A + i \Delta B) \eqcomma
      \label{eq:amp_shift}
  \\
  E & \to E + i \lambda \Delta \Sigma \eqcomma
      \label{eq:energy_shift}
\end{align}
and the correlation function at large times is given by
\begin{equation}
  G_\pm(\lambda, t)
  \atlarget
  \left[
    A \pm \lambda(\Delta A + i\Delta B)
  \right]
  e^{- \left[ E \pm i \lambda \Delta \Sigma \right] t}
  \eqstop
  \label{eq:corr_mod}
\end{equation}
(Recall that changing the spin projection corresponds to flipping the
sign of $\lambda$, as discussed in \sec{fh_methods}).
The quantity of interest is the shift in the phase, $\Delta \Sigma$.
To extract this value, we form the following ratio of real and
imaginary parts of spin-up and down projections,
\begin{align}
  R(\lambda, t)
  =
  \;
  &
  \frac
  {\Im \left[ G_-(\lambda,t) - G_+(\lambda,t) \right] }
  {\Re \left[ G_-(\lambda,t) + G_+(\lambda,t) \right] }
  \notag
  \\
  \atlarget
  \;
  &
  \frac
  {
    \sin \left( \lambda \Delta \Sigma t \right)
    - \lambda \frac{\Delta B}{A} \cos
    \left(\lambda \Delta \Sigma t \right)
  }
  {
    \cos \left( \lambda \Delta \Sigma t \right)
    + \lambda \frac{\Delta B}{A}
    \sin \left( \lambda \Delta \Sigma t \right)
  }
  \eqstop
  \label{eq:ratio}
\end{align}
Note that the form of this ratio does not change if we include second
order terms in \eq{amp_shift} and \eq{energy_shift}. For the operator
included in \eq{lag_mod}, we find that the second order shift
in the energy is purely real, and it can be shown that only the
factor $\frac{\Delta B}{A}$ will change.
Hence, corrections to these calculations do not appear until $\order{\lambda^3}$.

The ratio in \eq{ratio} is what we fit in our analysis.
To determine ground state saturation of this quantity,
we observe that, provided $t \ll \frac{1}{|\lambda \Delta \Sigma|}$,
the behaviour of the ratio is approximately linear in $t$.
\begin{equation}
  R(\lambda, t)
  =
  \lambda \Delta \Sigma t
  - \lambda \frac{\Delta B}{A}
  + \order{\lambda^3}
  \quad
  ,
  \quad
  a \ll t \ll \frac{1}{|\lambda \Delta \Sigma|}
  \eqstop
\end{equation}
Previous determinations of $\Delta \Sigma$
\cite{Babich:2010at, QCDSF:2011aa, Engelhardt:2012gd, Abdel-Rehim:2013wlz, Deka:2013zha}
suggest that
we should expect $|\Delta \Sigma| \approx 0.1$, and hence
for the largest value of $\lambda$ realised on our ensembles, $a\lambda = 0.05$,
this linear approximation will hold for times $\frac{t}{a} \ll
200$.
With this, we are able to introduce an `effective phase shift',
\begin{equation}
  \phi_\text{eff.}
  =
  \frac{1}{a}
  \left[
    \mathcal{R}(\lambda, t+a)
    -
    \mathcal{R}(\lambda, t)
  \right]
\end{equation}
which in the regime discussed has the behaviour
\begin{equation}
  \phi_\text{eff.}
  =
  \lambda \Delta \Sigma
  \quad
  ,
  \quad
  a \ll t \ll \frac{1}{|\lambda \Delta \Sigma|}
  \eqstop
  \label{eq:eff_phase}
\end{equation}

\section{Results}
\label{sec:results}

\begin{figure}
  \centering
  \includegraphics[width=\columnwidth]{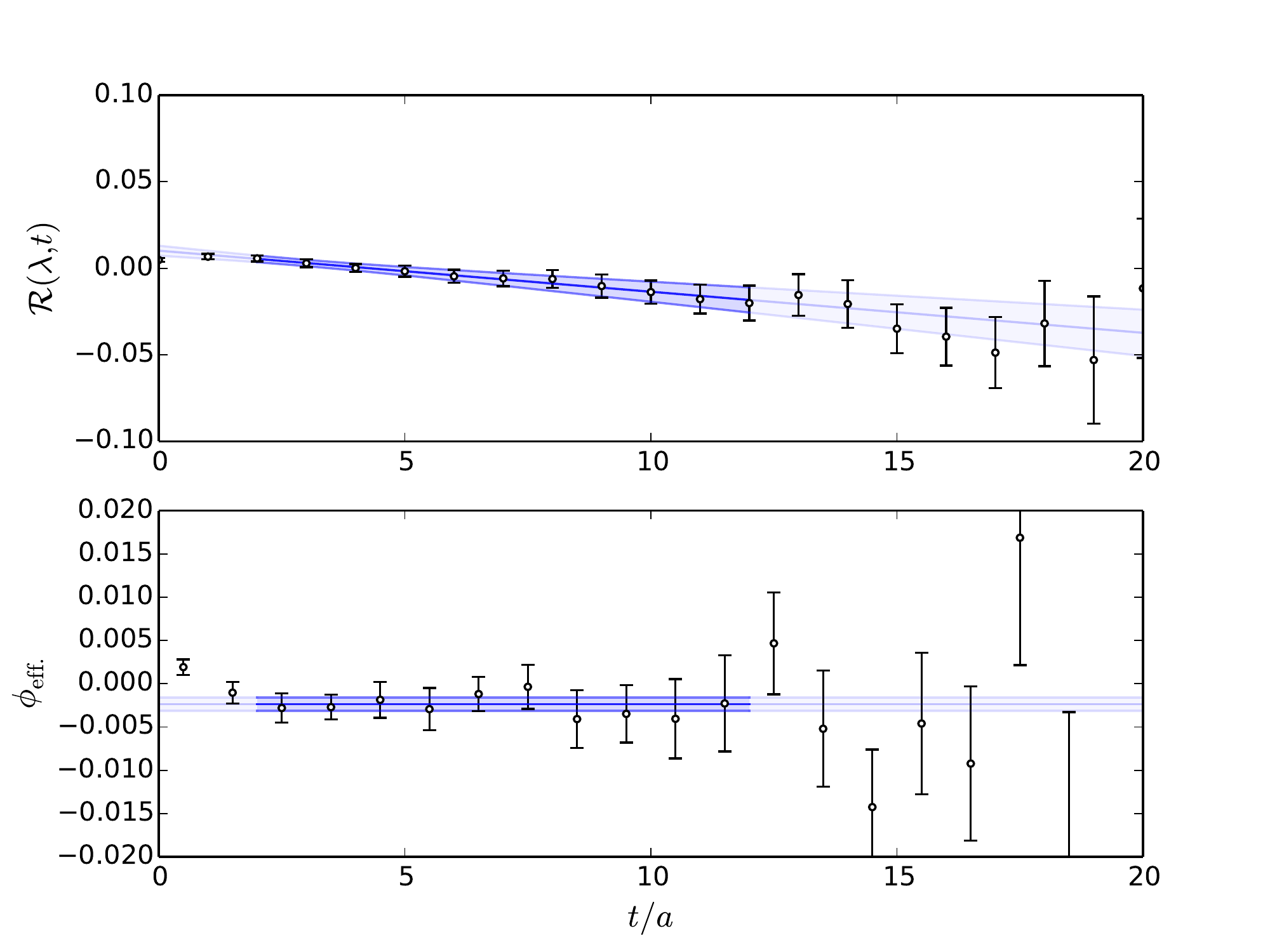}
  \caption{
    Plots of the ratio in \eq{ratio} and the effective phase shift
    defined in \eq{eff_phase} for $\lambda=0.03$,
    $m_\pi \approx 470$ MeV. The fitting window (shown in darker blue)
    was between time slices 2 and 12.
    The errors shown are from a bootstrap analysis of the results, as
    are the errors on the displayed fits.}
  \label{fig:eff_phase}
\end{figure}
\fig{eff_phase} shows an example plot of the ratio in \eq{ratio} and
the corresponding effective phase defined in \eq{eff_phase} for
$\lambda = 0.03$.  We observe a clear plateau in the effective phase
for the illustrated fitting region, and corresponding linear behaviour
in the ratio.  As an aside, the fit indicates a clearly non-zero value
for the $t = 0$ intercept, confirming that there is a small but
statistically significant imaginary shift in the wavefunction overlap
factors (given in \eq{amp_shift}) for this value of $\lambda$.

\begin{figure}
  \centering
  \includegraphics[width=\columnwidth]{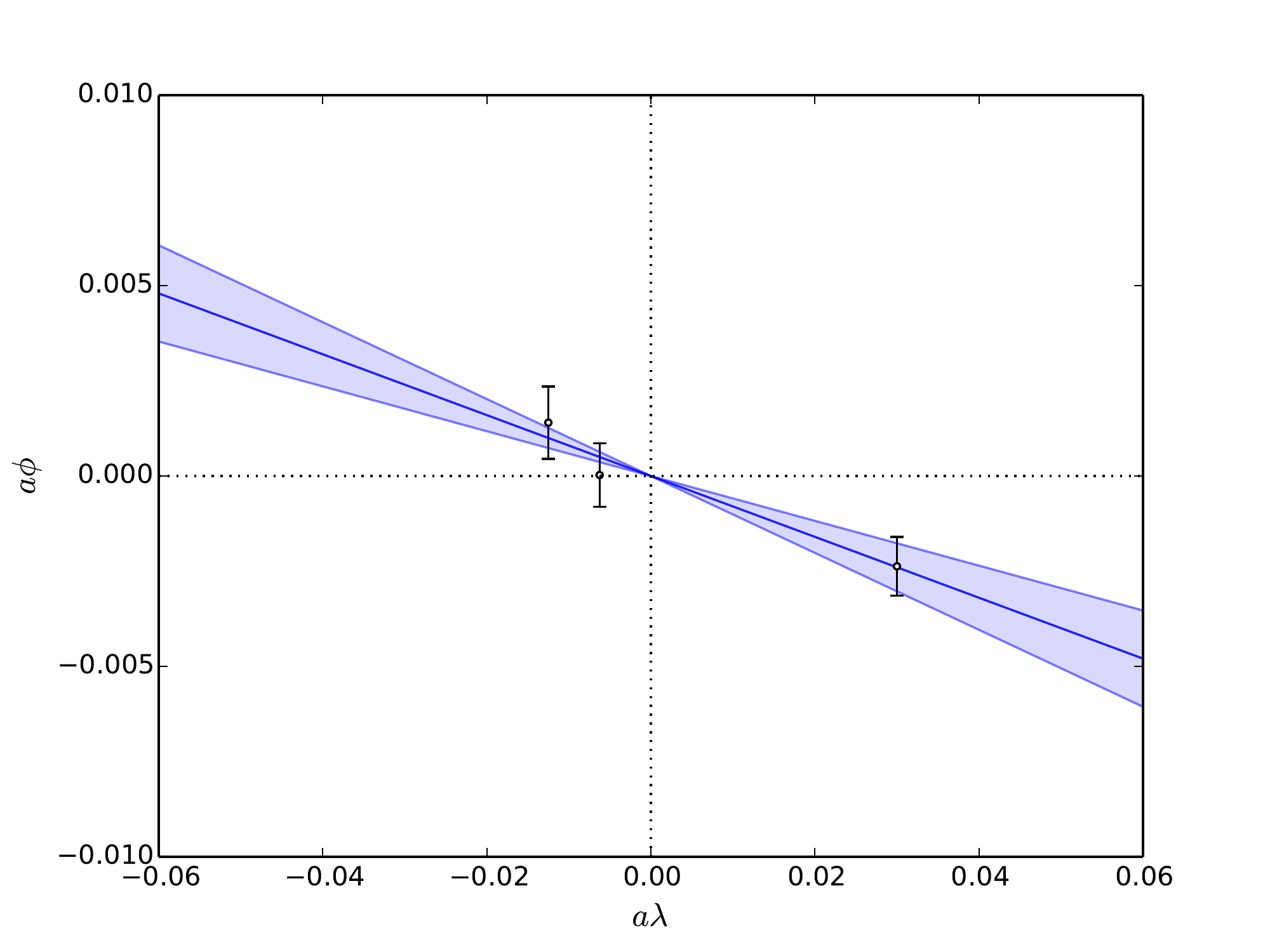}
  \caption{
    Phase shift as a function of $\lambda$ for $m_\pi \approx 470$ MeV.}
  \label{fig:fh_sym}
\end{figure}
Repeating this procedure for each value of $\lambda$ and extracting
the phase shifts, we are able to calculate the linear shift with
respect to $\lambda$, illustrated in \fig{fh_sym}.  From \eq{fh_rel}
we know that this shift is directly proportional to the disconnected
contribution to $\Delta \Sigma$.  Since there is no phase shift in the
zero-field limit, we have used a single-parameter slope fit to extract
the linear shift.  This analysis is repeated at the lighter pion mass.
\tab{ensembles} includes the calculated phase shift for each value of
$\lambda$ on the ensembles generated, and results of the described
analyses are summarised in \tab{results}.  Using the methods outlined
in \cite{Chambers:2014qaa}, we have also calculated the individual
connected contributions to $\Delta \Sigma$ on these ensembles, and
hence are able to calculate the total (connected and disconnected)
value of $\Delta \Sigma$.

\begin{table*}
  \begin{tabular}{c | c c c c | c c }
    \hline
    \hline
    $(\kappa_l, \kappa_s)$ & $\Delta u^\text{latt.}$ & $\Delta d^\text{latt.}$ & $\Delta \Sigma_{\text{disc.}}^\text{latt.}$
    & $\Delta \Sigma^\text{latt.}$
    & $\Delta \Sigma_\text{disc.}^{\overline{\text{MS}}(2 \text{ GeV}^2)}$
    & $\Delta \Sigma^{\overline{\text{MS}}(2 \text{ GeV}^2)}$ \\
    \hline
    $(0.120900, 0.120900)$ & 1.001(7)  & $-0.310(5)$ & $-0.079(21)$ & 0.612(24) & $-0.055(18)$ & 0.530(21) \\
    $(0.121095, 0.120512)$ & 1.004(10) & $-0.319(6)$ &  0.014(16) & 0.699(25) &  0.026(14) & 0.605(21) \\
    \hline
    \hline
  \end{tabular}
  \caption{Table of results at each pion mass for the individual quark
    axial charges and the disconnected and full (connected plus
    disconnected) contribution to the total quark spin. The quantities reported with the ``latt.'' superscript are unrenormalised. The final two columns report our renormalised results based upon Eqs.~(\ref{eq:renorm}) through (\ref{eq:renormDisc}).}
  \label{tab:results}
\end{table*}

At the lighter mass, we find a result consistent with zero
for $\Delta \Sigma_\text{disc.}$.
This unusual result may be the result of a couple of different factors.
The $\lambda$ values chosen may simply be too small, and the phase
shift is not able to rise above the background correlator noise.
\begin{figure}
  \centering
  \includegraphics[width=\columnwidth]{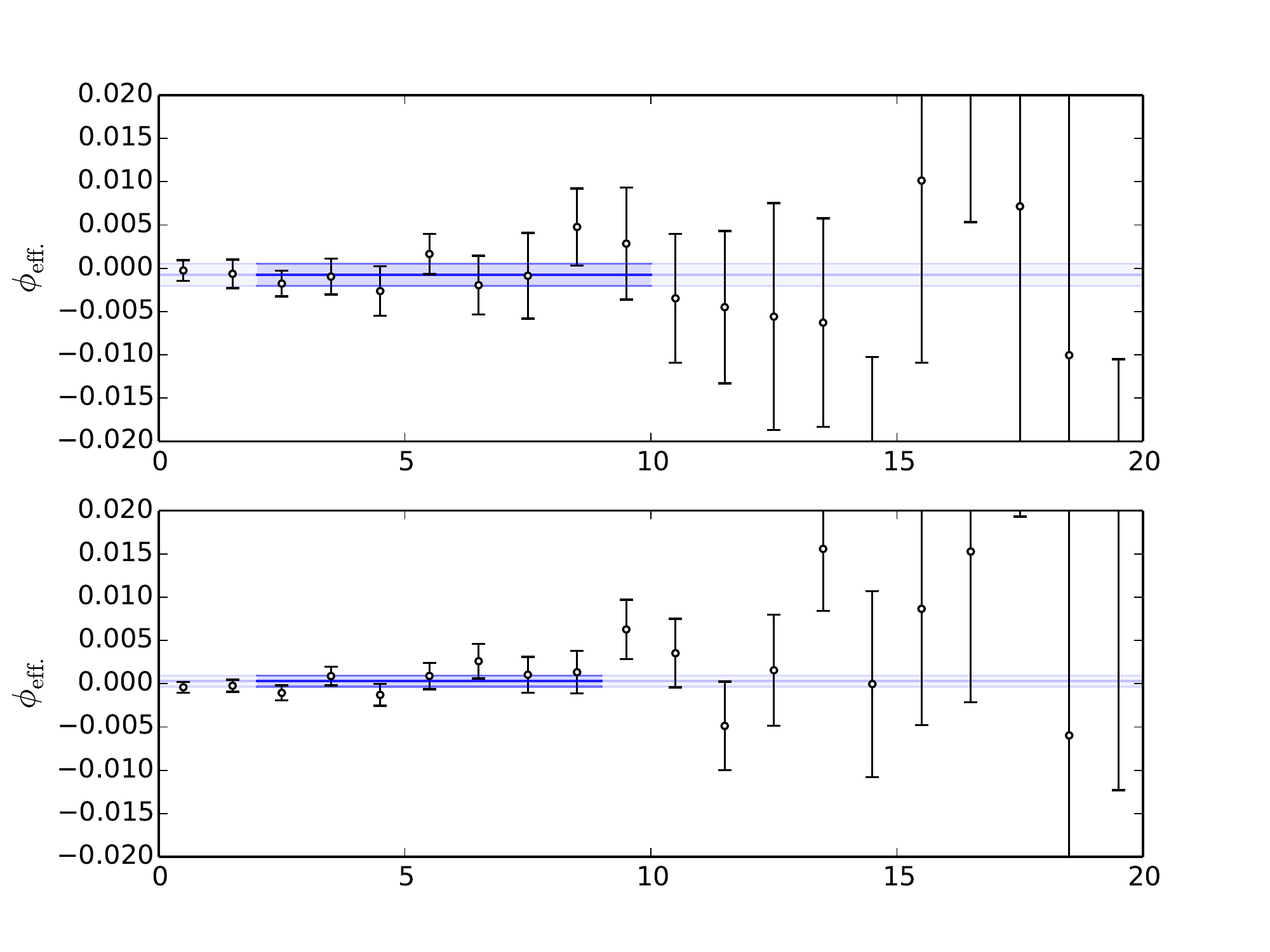}
  \caption{
  Effective phase plots for $\lambda=-0.025, 0.05$ respectively
  at $m_\pi \approx 310$ MeV.
  The results in the second plot have greater statistics by
  a factor of 4.
  Note that the sign of the fitted value is highly dependent
on the fit window.}
  \label{fig:light_ep}
\end{figure}
\fig{light_ep} show effective phase plots for the two values of
$\lambda$ realised at this lighter quark mass, and show that there is
no clearly identifiable plateau at these statistics.

Alternatively, there may be a sign change in either the light or
strange contribution to $\Delta \Sigma$ at some mass between $m_\pi =
310 -470$ MeV.
This is unlikely, however, as previous results at similar masses have shown
a significant $\Delta s$ contribution, which would require
the light quark contribution to have a strong quark mass dependence.

From \cite{Chambers:2014pea} we have both non-singlet and singlet
renormalisation factors for the axial current at the SU(3) symmetric
point,
\begin{align}
  Z_{A,\text{NS}}^{\overline{\text{MS}}(2 \text{ GeV}^2)} & = 0.8458(8)
  \eqcomma   \label{eq:renorm} \\
  Z_{A,\text{S}}^{\overline{\text{MS}}(2 \text{ GeV}^2)} & = 0.8662(34)
  \eqstop
\end{align}
Further calculations at additional quark masses are required
to perform a chiral extrapolation of these quantities, however the pion
mass dependence of these factors is expected to be mild based on the
non-singlet calculation of Ref.~\cite{Constantinou:2014fka}.

To obtain the renormalised total spin contribution we use the singlet
renormalisation factor:
\begin{equation}
\Delta \Sigma^{\overline{\text{MS}}}
=Z_{A,\text{S}}^{\overline{\text{MS}}}\Delta \Sigma^{\text{latt.}}.
\end{equation}
For the purely disconnected quantity, we include the mixing of the
connected and disconnected contributions under renormalisation:
\begin{equation}
  \Delta \Sigma_\text{disc.}^{\overline{\text{MS}}}
  =
  Z_{A,\text{S}}^{\overline{\text{MS}}} \Delta \Sigma_\text{disc.}^\text{latt.}
  +
  \left(
    Z_{A,\text{S}}^{\overline{\text{MS}}}
    -
    Z_{A,\text{NS}}^{\overline{\text{MS}}}
  \right)
  \Delta \Sigma_\text{conn.}^\text{latt.}
  \eqstop
\label{eq:renormDisc}
\end{equation}
Using the renormalisation factors from the SU(3) symmetric point, we
quote our $\overline{\text{MS}}$ results in the final two columns of
\tab{results}.

Finally, since at the SU(3) symmetric point all quarks contribute
exactly the same amount to $\Delta \Sigma_{\text{disc.}}^{\overline{\text{MS}}}$, then at this point we
can determine $\Delta s$
\begin{equation}
\Delta s^{\overline{\text{MS}}}(m_\pi=465\text{MeV}) =\frac13 \Delta\Sigma_{\text{disc.}}^{\overline{\text{MS}}}= -0.018(6)\ .
\end{equation}

\section{Concluding remarks}

Culminating in the results of \tab{results},
we have shown how the Feynman-Hellmann theorem may be applied to perform
full calculations of hadronic matrix elements.

Extensions of these particular calculations include higher-statistics
simulations, particularly at the lighter pion mass, and the generation
of ensembles at additional pion masses to identify the quark mass
dependence of $\Delta \Sigma$. Further analysis of the existing data
should allow for the extraction of disconnected quark spin
contributions for the other octet baryons and the vector mesons.

The FH technique demonstrated here can be easily adapted to a variety
of other disconnected quantities, such as the gluonic contribution to
angular momentum, which would otherwise be rather challenging using
conventional approaches.

\section*{Acknowledgements}

The numerical configuration generation was performed using the BQCD
lattice QCD program, \cite{Nakamura:2010qh}, on the IBM BlueGeneQ
using DIRAC 2 resources (EPCC, Edinburgh, UK), the BlueGene P and Q at
NIC (J\"ulich, Germany) and the Cray XC30 at HLRN
(Berlin--Hannover, Germany).
Some of the simulations were undertaken using resources awarded at the
NCI National Facility in Canberra, Australia, and the iVEC facilities
at the Pawsey Supercomputing Centre. These resources are provided
through the National Computational Merit Allocation Scheme and the
University of Adelaide Partner Share supported by the Australian
Government.
The BlueGene codes were optimised using Bagel \cite{Boyle:2009vp}.
The Chroma software library \cite{Edwards:2004sx}, was used in the
data analysis.
This investigation has been supported
by the Australian Research Council under grants
FT120100821, FT100100005 and DP140103067 (RDY and JMZ).
HP was supported by DFG grant SCHI 422/10-1.

\bibliography{ref}

\end{document}